\documentclass[12pt,preprint]{aastex}

\usepackage[dvips]{color}

\def\bracket#1{\left\langle\mbox{$#1$}\right\rangle}
\def\e{\mathrm{e}}

\slugcomment{Draft version \today}
\shorttitle{IMPACT OF ROTATION ON NEUTRINO EMISSION FROM POP III}
\shortauthors{SUWA ET AL.}

\begin{document}

\title{ Impact of Rotation on Neutrino Emission \\
  and Relic Neutrino Background from Population III Stars}

\author{Yudai Suwa\altaffilmark{1}, Tomoya Takiwaki\altaffilmark{2}, Kei Kotake\altaffilmark{3,4},
  and Katsuhiko Sato\altaffilmark{1,2,5}} 

\altaffiltext{1}{Department of Physics, School of Science, the University of Tokyo, 7-3-1 Hongo,
  Bunkyo-ku, Tokyo 113-0033, Japan} 
\altaffiltext{2}{Research Center for the Early Universe, School of Science, the University of
  Tokyo,7-3-1 Hongo, Bunkyo-ku, Tokyo 113-0033, Japan}
\altaffiltext{3}{Division of Theoretical Astronomy, National Astronomical Observatory of Japan,
  Mitaka, Tokyo 181-8588, Japan}
\altaffiltext{4}{Center for Computational Astrophysics, National Astronomical Observatory of Japan,
  Mitaka, Tokyo 181-8588, Japan}
\altaffiltext{5}{The Institute for the Physics and Mathematics of the Universe, the University of
  Tokyo, Kashiwa, Chiba, 277-8568, Japan}

\email{suwa@utap.phys.s.u-tokyo.ac.jp}


\begin{abstract} 
    We study the effects of rotation on the neutrino emission from Population III (Pop III) stars by
    performing a series of two-dimensional rotational collapse simulations of Pop III stellar cores.
    Our results show that rotation enhances the neutrino luminosities and the average energies of
    emitted neutrinos. This is because the thermalized inner core, which is the dominant neutrino
    source from Pop III stars, can be enlarged, due to rotational flattening, enough to extend the
    inner core outside the neutrinospheres.  This is in sharp contrast to the case of spherical
    collapse, in which the case of inner core shrinks deeper inside the neutrinospheres before black
    hole formation, which hinders the efficient neutrino emission.  In the case of rotational
    core-collapse, the emitted neutrino energies are found to become larger in the vicinity near the
    pole than the ones near the equatorial plane.  These factors make the emergent neutrino spectrum
    broader and harder than the spherical collapse case. By computing the overall neutrino signals
    produced by the ensemble of individual rotating Pop III stars, we find that the amplitudes of
    the relic neutrinos, depending on their star formation rates, can dominate over the
    contributions from ordinary core-collapse supernovae below a few MeV.  A detection of this
    signal could be an important tool to probe star formation history in the early universe.
\end{abstract}

\keywords{hydrodynamics --- methods: numerical --- neutrinos ---  stars: rotation }

\section{Introduction}

One of the most important goals in modern cosmology is to understand how the first stars formed at
the end of the dark ages, and how they transformed the initially simple, homogeneous universe into a
state of ever increasing complexity \citep[e.g.,][]{bark01,brom04,loeb06}.  The first stars, so
called Population III (Pop III), are predicted to have been predominantly very massive with
$M\gtrsim 100M_\odot$ \citep[e.g.,][]{naka01,abel02, brom02b,yosh06,oshe07}.  During their
evolutions, the central core is thought to experience an electron-positron pair creation instability
after carbon burning, which reduces the thermal energy of the core and eventually triggers
gravitational core-collapse.  For stellar masses less than $\sim 260 M_\odot$, rapid nuclear burning
releases a large amount of energy sufficient to entirely disrupt the star as pair-instability
supernovae.  More massive stars, which also encounter pair-instability, are so tightly bound that
the fusion of oxygen is unable to reverse infall.  Such stars are thought to collapse into black
holes (BHs) \citep{bond84,fryer01}, which we investigate in this paper.  Information about the
formation and evolution of the Pop III stars might be observationally obtained from reionization
\citep{alva06}, the infrared background \citep{dwek05}, nucleosynthesis yields
\citep{heger02,iwam05}, gamma-ray bursts \citep{schn02,brom06}, and gravitational waves
\citep{buon05,suwa07b}.  However at present, it is still unclear whether or not one can obtain clear
information about Pop III stars directly from observations.

Recently, neutrinos, which should have been emitted in large numbers at the phase of core-collapse,
are expected to be new eyes to unveil the mysteries of the first stars.  Since the direct detection
is unlikely because of their large distances, a diffuse background as relic neutrinos could
potentially be observable.  In an analogous context, the relic neutrino background from ordinary
core-collapse supernovae have been studied elaborately \citep{bisn84,tota96,kapl00,ando03,ando04}.
Since the physics involved in relic neutrinos covers a very wide range of topics, e.g., cosmic star
formation rate and the collapse dynamics, detecting the relic neutrinos or even setting limits on
their flux are expected to give us quite useful and unique implications not only for the
understanding of the collapse physics itself, but also for various fields of astrophysics and
cosmology.
 
The detectability of the relic neutrinos from Pop III stars has been discussed by several papers.
\cite{iocco05} estimated the amplitudes of the relic neutrinos produced during the nuclear burning
phases as well as the core collapse phase, assuming a baryonic fraction of Pop III stars of
$10^{-3}$ and monochromatic progenitors with a mass of $300 M_\odot$.  \cite{daig05} investigated
the relic neutrino background produced by an early burst of Pop III stars taking into account a
``normal mode'' of the star formation at low redshift.  The framework of hierarchical structure
formation, on which their computation relies, is based on the cosmic star formation histories, which
are constrained by the observed star formation rate at redshift $z\lesssim 6$, the observed chemical
abundances in damped Lyman $\alpha$ absorbers and in the IGM, and allow for an early reionization of
the Universe at $z\sim 10-20$.  Regardless of the differences in the treatment of the star formation
histories, these studies concluded that the typical energy of the relic neutrino background is
lowered as low as MeV or sub-MeV due to the cosmological redshift so that the detection is out of
the question with presently known experimental techniques. However it should be noted that in their
studies, the neutrino flux and energy emitted from the Pop III stars are set by hand due to the lack
of Pop III star collapse simulations.

For more precise estimates of relic neutrinos from Pop III stars, we have to perform the
hydrodynamical simulations.  So far, there have been only a few simulations studying the
gravitational collapse of Pop III stars with BH formation including appropriate microphysical
treatment \citep{fryer01, naka06,suwa07a}.  \cite{naka06} performed a spherically symmetric general
relativistic simulation in a wide range of masses ($300-10000 M_\odot$), in which the
state-of-the-art neutrino physics are taken into account.  Their detailed calculations revealed the
properties of the emergent neutrino spectrum, and based on that, they estimated the relic neutrino
background and found that the detection for the spherically collapsing Pop III stars is difficult
for the currently operating detectors.  Meanwhile current wisdom tells us that very metal poor stars
lose very small amounts of mass and angular momentum through radiatively driven stellar winds so
that Pop III stars may collapse with large angular momentum \citep{hege03a}.  So stellar rotation
might vary the spectrum of emitted neutrinos and affect its potential observability, which is the
motivation of this study.  \cite{fryer01} investigated the collapse of a rotating Pop III star of
mass $300 M_\odot$ with gray neutrino transport simulations and discussed effects of rotation on the
emitted luminosities.  They, however, investigated only one rotational model. Considering the
uncertainty of the angular momentum distributions of the Pop III stars, a more broad study is
required.  More recently, we implemented a series of two-dimensional magnetorotational core-collapse
simulations of Population III stars \citep{suwa07a}, in which our interest was placed on the MHD
effects on the formation of the jets.

In this study, we focus on the effects of rotation on the neutrino emission during the core-collapse
phase of Pop III stars. For this purpose, we carry out two-dimensional simulations of rotational
core-collapse of Pop III stars, changing the initial strength of the angular momentum in a
parametric manner.  By so doing, we systematically investigate the dependence of the luminosity, the
average energy, and the spectrum of neutrinos upon the rotation strength.  We also pay attention to
the anisotropy of the emergent neutrino energies, that is, the neutrino emission at polar direction
and in the equatorial plane.  Using the spectrum of single Pop III stars, we calculate the number
flux of the relic neutrino background by summing up the contribution from individual Pop III stars
and discuss their detectability.

This paper is organized as follows: In the next section, we briefly introduce our numerical methods
and the initial models.  In \S 3 and \S4, we present the numerical results. We start with the
outline of the dynamics and neutrino emission in the case of spherical collapse (\S\ref{sec:sph})
and then move on to discuss the effects of rotation on the dynamics and the neutrino luminosity
(\S\ref{sec:lumi}), the neutrino energies (\S\ref{sec:ene}) and the spectrum (\S\ref{sec:spe}).  \S
5 is devoted to summary and discussion.

\section{Numerical Techniques and Initial Models}

As for the hydro solver, we employ the ZEUS-2D code \citep{ston92} as a base and added major changes
to include the microphysics appropriate for the simulation of Pop III stars.  First we have added an
equation for the electron fraction in order to treat electron and positron captures and have
approximated the neutrino transport by the so-called leakage scheme
\citep{ruff96,kota03b,ross03,taki07}.  We consider three neutrino flavours: electron neutrinos,
$\nu_e$, electron anti-neutrinos, $\bar\nu_e$, and the heavy-lepton neutrinos, $\nu_\mu,
\bar\nu_\mu, \nu_\tau, \bar\nu_\tau$, which are collectively referred to as $\nu_X$.  Reactions of
$\nu_X$, pair, photo, and plasma processes are included using the rates by \cite{itoh89}.  As for
the equation of state, we have incorporated the tabulated one based on relativistic mean field
theory \citep[see,][]{shen98}.  In our 2D calculations, axial symmetry and reflection symmetry
across the equatorial plane are assumed.  Spherical coordinates $(r,\theta)$ are employed with
logarithmic zoning in the radial direction and linear zoning in $\theta$.  One quadrant of the
meridian section is covered with 300 ($r$)$\times$ 30 ($\theta$) mesh points.  We also calculated
some models with 60 angular mesh points, however, there was no significant difference.  Therefore,
we report only the results obtained from the models with 30 angular mesh points.  General
relativistic gravity is taken into account approximately by an ``effective relativistic potential''
according to \cite{mare06}, which is newly added in our code.

In this paper, we set the mass of the Pop III star to be $300 M_{\odot}$.  This is consistent with
the recent simulations of the star-formation phenomena in a metal free environment, providing an
initial mass function peaked at masses $100 - 300 M_{\odot}$ \cite[see, e.g.,][]{naka01}.  We choose
the value because we do not treat the nuclear-powered pair instability supernovae ($M \lesssim 260
M_{\odot}$) and also to facilitate the comparison with the previous study, which employed the same
stellar mass \citep{fryer01}.

We start the collapse simulations with a $180M_\odot$ core for the $300M_\odot$ star. The core,
which is the initial condition of our simulations, is produced in the following way. According to
the prescription in \cite{bond84}, we set the polytropic index of the core to $n=3$ and assume that
the core is isentropic with $\sim 10 k_B$ per nucleon \citep{fryer01} with a constant electron
fraction of $Y_e=0.5$. We adopt a central density of $5\times10^6~{\rm g}~{\rm cm^{-3}}$ at which
the temperature of the central regions become high enough to photodisintegrate the iron ($\sim
5\times10^9$K), thus initiating the collapse.  Given the central density, the distribution of
electron fraction, and the entropy, we numerically construct the hydrostatic structures of the core.

Since we know little about the angular momentum distribution in the cores of Pop III stars, we
assume the shellular rotation follows the initial rotation law:
\begin{equation}
    \Omega(r)=\Omega_0\frac{r_0^2}{r^2+r_0^2},
\end{equation}
where $\Omega(r)$ is an angular velocity, $r$ is a radius, and $\Omega_0$ and $r_0$ are model
parameters, which determines a total angular momentum and the degree of differential rotation,
respectively. In this paper, we fixed the value of $r_0$ as $10^9$ cm.  Since the radius of the
outer edge of the core is taken to be as large as $3.5 \times 10^9$ cm, the above profile represents
a mildly differentially rotating core.  Changing the initial rotational energies by varying the
values of $\Omega_0$, we compute 8 models, namely, one spherical and 7 rotational models. By so
doing, we hope to see clearly how the dynamics and the neutrino emission deviate from those in the
spherical case.  We change the initial values of $T/|W|$ from 0.01 to 2 $\%$, where $T/|W|$
represents the ratio of the rotational to the gravitational energy.  We take the model with $T/|W|$
0.5 $\%$ as the canonical rotating model. This is because the rotation rate is taken from the
stellar evolution calculations of extremely rapidly rotating cores of massive stars \citep{hege00}
pushed by the study that Pop III stars could rotate rapidly due to insufficient mass-loss in the
main sequence stage \citep{hege03a}.

The shape of the neutrino spheres is important for analyzing the properties of neutrino emission.
Following the common practice, we define the neutrinosphere, $R_\nu$, satisfies the condition
\begin{equation}
    \int^\infty_{R_{\nu}}\frac{dr}{\lambda}=\frac{2}{3},
\end{equation}
where $\lambda$ is the mean free path of neutrinos. We perform the integration for each angular bin
to obtain $R_{\nu}(\theta)$. It should be noted that the notion of neutrinosphere is rather
ambiguous.  Neutrino reactions are highly energy dependent, and so the neutrinosphere should be as
well. In theory, we should distinguish the last scattering surface from the surface of last energy
exchange \citep{jank95}. However, we stick to this approximate notion here for simplicity.

\section{Spherical Collapse} \label{sec:sph}

\begin{figure*}[p]
    \centering
    \includegraphics[width=\linewidth]{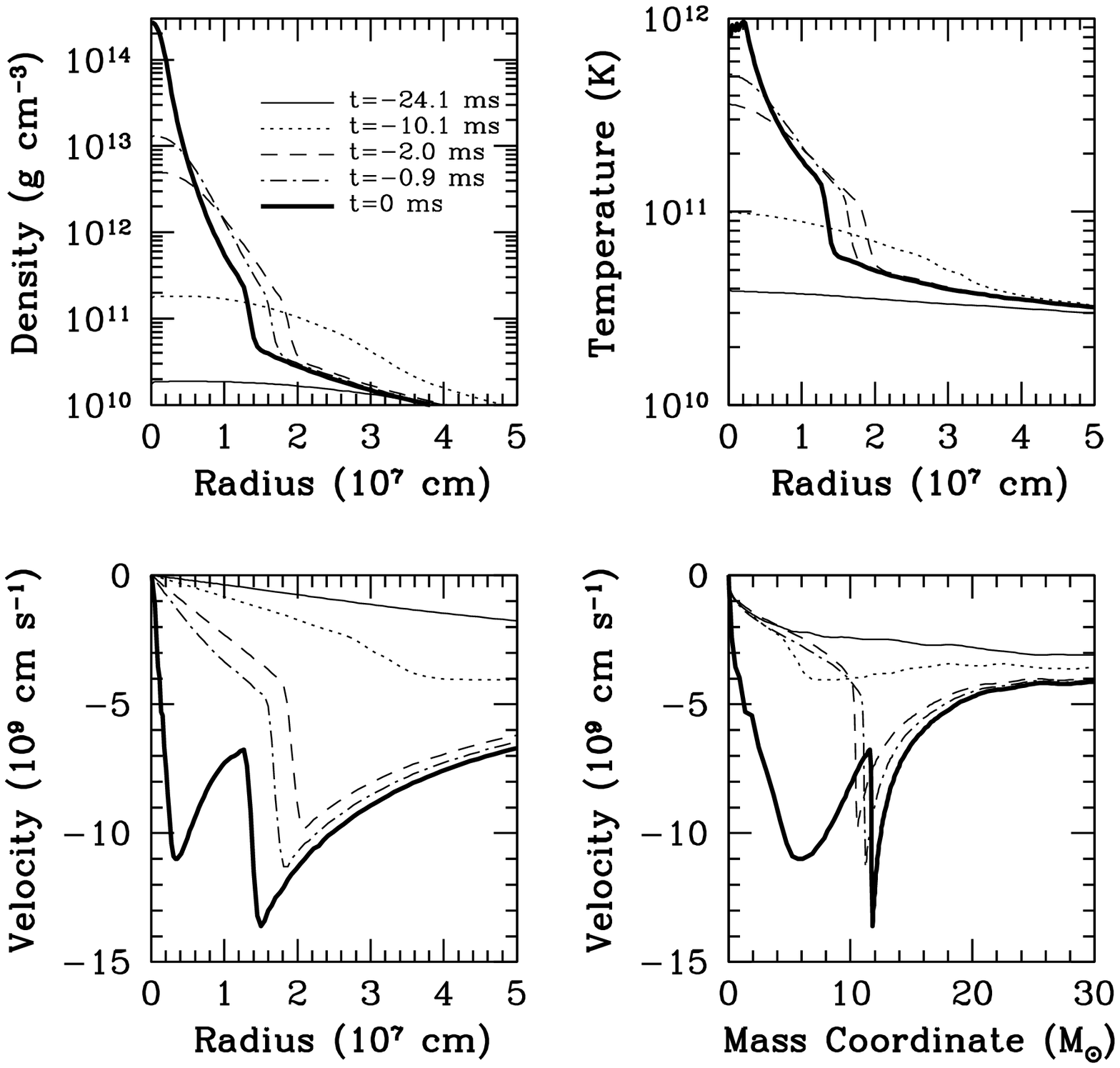}
    \caption{ Time evolution of density (top left), temperature (top right), and radial velocity
      (bottom left) for non-rotating model as a function of radius in $10^7$ cm. In addition, the
      radial velocity as a function of mass coordinate in $M_\odot$ (bottom panels) is presented for
      comparison. The thin-solid, dotted, dashed, dot-dashed, and thick solid lines are snapshots at
      24.1, 10.1, 2.0, 0.9, and 0 ms before black hole formation, respectively.}
    \label{fig:den}
\end{figure*}

We first outline collapse dynamics in the case of spherical symmetry.  In Figure \ref{fig:den}, we
show the time evolution of density (top left), temperature (top right), and radial velocity (bottom
left) as a function of radius.  Also shown is the radial velocity as a function of the mass
coordinate (bottom right).  We show the results for five time slices before BH formation (see figure
for detail).  As indicated in this figure, the core collapse of a very massive star can be divided
into two phases; the infall phase ($-24.1$ and $-10.1$ ms) and the accretion phase (from $-2.0$ to
$0$ ms).

The infall phase sets in due to the gravitational instability induced by the photodisintegration of
iron and the neutronisation that have the effect of reducing the pressure support at the core of the
star, triggering a rapid collapse of the core just as an ordinary supernovae progenitor.  However,
there are several important differences between the structures of very massive stars and ordinary
supernovae. As noted by previous works \citep[e.g.,][]{bond84,fryer01,naka07}, the entropy in the
more massive core is larger and thus favors more complete photodisintegration of heavy elements and
$\alpha$-particles so that the source of instability is reduced. This leads a thermally stabilized
core
\footnote{ The stability of the core depends on the adiabatic index. If the average adiabatic index
  is less than $4/3$, the star is unstable \citep{shap83}. In the case of very massive stars, the
  pressure is dominated by radiation pressure, whose adiabatic index is $4/3$. Since there are
  additional components such as the non-relativistic nucleon, whose adiabatic index is $5/3$, the
  {\it mean} adiabatic index is between $4/3$ and $5/3$ so that the core is mildly stabilized.  }
unlike ordinary supernovae, whose core is stabilized by nuclear forces and neutron degeneracy
pressure. The core is divided into two distinct regions: the subsonic, ``homologous'' inner core and
the supersonic outer core. The inner core becomes thermalized and stabilized.

\begin{figure}[p]
    \centering
    \includegraphics[width=\linewidth]{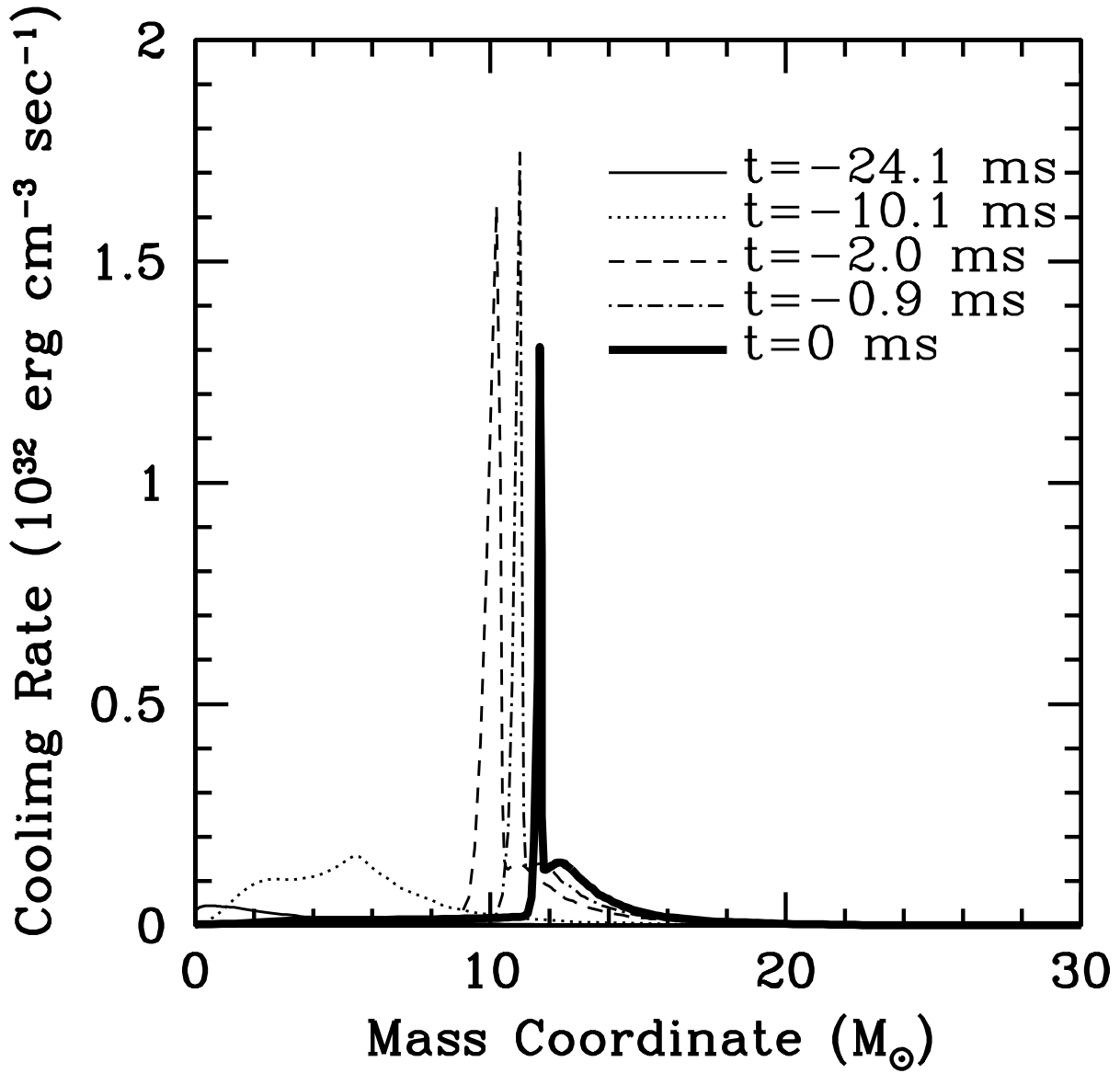}
    \caption{ Snapshots of total neutrino cooling rate $Q_\nu$ in $10^{32}$ erg s$^{-1}$ cm$^{-3}$
      as a function of mass coordinate in $M_\odot$, for the non-rotating model, showing that the
      most efficient cooling appears just behind the accretion shock. Times are relative to black
      hole formation time. Compare this figure with Figure \ref{fig:den}, which shows the
      hydrodynamical evolution for this same model.}
    \label{fig:emis}
\end{figure}

In the accretion phase, material in the outer core accretes onto the inner core after the thermal
stabilization of the inner core, leading to the increase in mass of the inner core. We here note
that the inner core continues to shrink after stabilization due to the accretion from the outer
core.  These features are seen in Figure \ref{fig:den}. In the top panels and bottom left panel, the
accretion shock can be seen at $\sim 2\times 10^7$ cm for $t=2.0$ ms and $\sim 1.5\times 10^7$ cm
for $0$ ms before BH formation. The bottom right panel shows the increasing mass of the inner core.
Because the outer core is out of sonic contact with its inner counterpart, an accretion shock is
formed between the two regions
\footnote{This accretion shock is similar to the stalled and hydrostatic shock front in the
  framework of ordinary core collapse supernovae. The difference is that there is no outgoing shock
  wave for very massive stars due to the softness of the equation of state inside the inner core.}.
It should be noted that the region just behind the accretion shock, that is the surface of the inner
core, is the main source of neutrinos.  This can be seen in Figure \ref{fig:emis}, which shows
snapshots of the cooling rate of neutrinos, $Q_\nu$, as a function of the mass coordinate. Comparing
with Figure \ref{fig:den}, we find that the maximum cooling rate is just behind the accretion shock.

In the case of very massive stars, the inner core is more massive than the maximum mass of neutron
stars so that these stars collapse to black holes directly.  As for the non-rotating model, the
initial mass of the BH is $\sim 5 M_\odot$.  This is significantly smaller than our previous result,
$20 M_\odot$ for the non-rotating model in \citet{suwa07a}, because of general relativistic effects,
newly implemented in this study. Although the core is stable in Newtonian gravity, the hot core is
not stable when general relativistic effects are considered because the thermal energy also
contributes to the gravitational potential.  In our case the temperature is so high that general
relativistic effects become important because the critical temperature, above which the core becomes
unstable, is $T_\mathrm{crit}\sim 10^{12} \mathrm{K} (10M_\odot/M)$, where $M$ is the core mass
\citep{shap83}.  Our results show the inner core mass is $\sim 10 M_\odot$ and the central
temperature is $\sim 10^{12}$ K (see the discontinuity of bottom right panel in Figure \ref{fig:den}
for core mass and top right panel for temperature).  This instability drives further collapse of the
inner part of the inner core, which can be seen in the radial velocity profile (see again bottom
panels in Figure \ref{fig:den}).  This feature is also seen in \citet{naka06}, which is a fully
general relativistic simulation (see Figure 2 in their result).  Although the BH is smaller than the
inner core at first, the inner core will be soon swallowed completely.  Here we note that following
\cite{fryer01,suwa07a}, BH formation is ascribed to the condition $\frac{6Gm(r)}{c^2}>r$, where
$c,G,m(r)$ are the speed of light, the gravitational constant, and the mass coordinate,
respectively.  After formation, we excise the region inside and then treat it as an absorbing
boundary. Then, we enlarge the boundary of the excited region to take into account the growth from
the mass infalling into the central region.  Although it is not accurate at all to refer to the
central region as a BH, we adhere to this simplification in order to follow and explore the dynamics
later on.

\begin{figure}[p]
    \centering
    \includegraphics[width=0.9\linewidth]{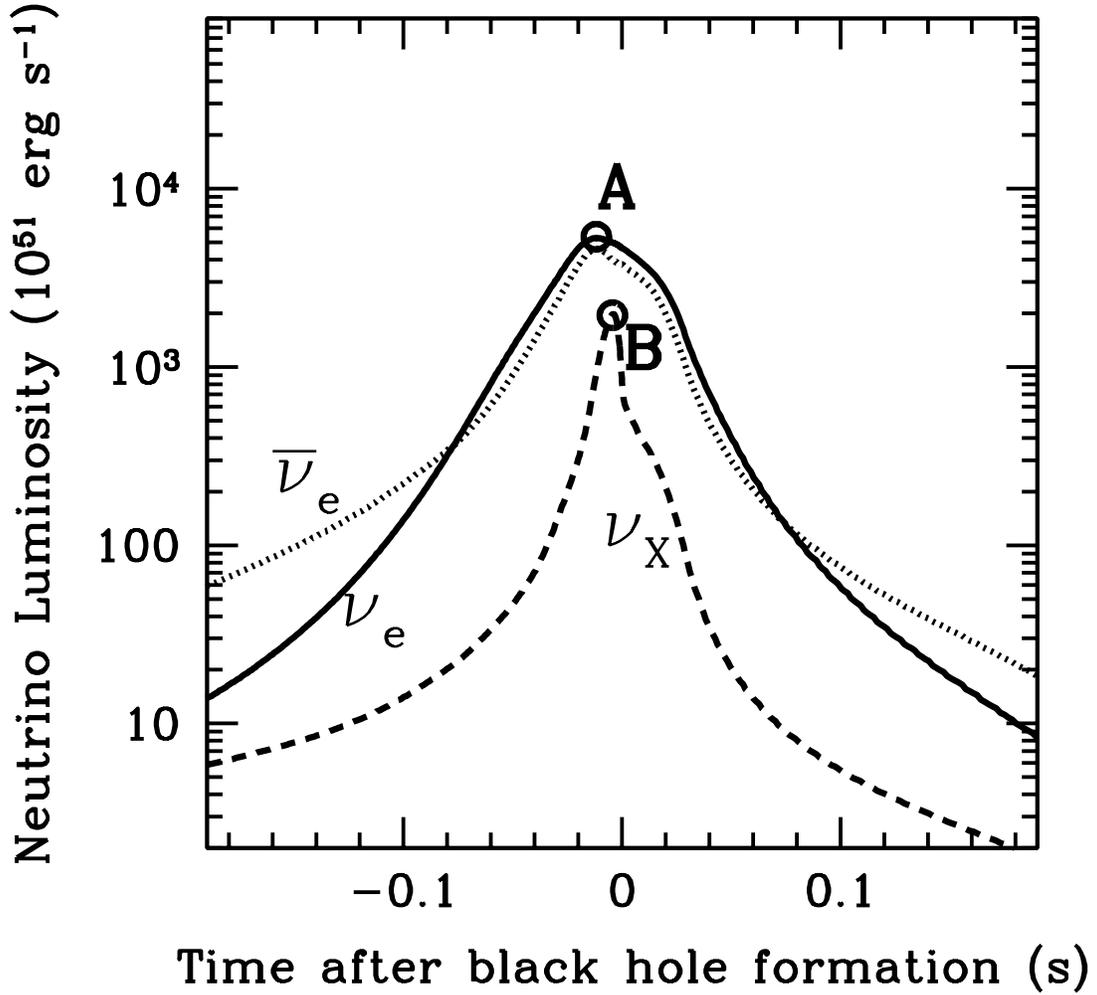}
    \caption{ Time evolution of neutrino luminosity $L_\nu$ in $10^{51}$ erg s$^{-1}$ for the
      non-rotating model. Solid, dotted and dashed lines represent $\nu_e$, $\bar\nu_e$ and $\nu_X$,
      respectively. Circles A and B correspond to the peak luminosity for $\nu_e$ and $\nu_X$.}
    \label{fig:lumi_sp}
\end{figure}

Next, we mention the neutrino emission and its relation to dynamics.  The time evolution of the
neutrino luminosity, $L_\nu \equiv \int Q_\nu dV$, is shown in Figure \ref{fig:lumi_sp}. As the
collapse proceeds, the density and temperature of the central both increase, increasing the neutrino
luminosity.  As the density increases during core collapse, the opacity rises and the neutrinosphere
is formed, in which neutrino emission-absorption equilibrium is achieved.  The relation of the radii
of the inner core surface, $R_\mathrm{ic}$, and neutrinosphere, $R_\nu$, is very important for the
luminosity of neutrinos because the inner core is the main source of neutrino emission as already
shown in Figure \ref{fig:emis}.  Roughly speaking, if $R_\mathrm{ic} \gtrsim R_\nu$, the emissivity
is the largest just behind the shock so that the luminosity is also large.  On the other hand, the
luminosity becomes small if $R_\nu \gtrsim R_\mathrm{ic}$ because the emitted neutrinos can not
escape directly and interact with matter.  In this case, the maximum value of the emissivity appears
between the inner core surface and the neutrinosphere. In Figure \ref{fig:ns} we show the evolution
of radii of the inner core, neutrinospheres, and black hole as a function of time. The inner core
surface goes down inside the neutrinospheres, which results in the peak in the evolution of
luminosity shown in Figure \ref{fig:lumi_sp} (see dots A and B in Figure \ref{fig:lumi_sp} and
\ref{fig:ns}).  In addition, after BH formation the luminosity decreases due to the absorption of
neutrino emitting matter.  Finally, the neutrinosphere gets swallowed by the BH and the emission of
neutrinos almost terminates.

\section{Rotational Collapse}

\subsection{Features of Neutrino Luminosity} \label{sec:lumi}

We now investigate how rotation affects the properties of the neutrino emission.  Peak luminosity
and total energy emitted by neutrinos for different values of the initial rotation rate,
$T/|W|_\mathrm{init}$, are shown in Table \ref{tab}.  Interestingly and maybe against the intuition,
it is found that the peak neutrino luminosity increases with the initial rotation rate (see Table
\ref{tab}).  We point out that this is because the inner core can partly extend outside the
neutrinosphere due to rotation so that $R_\mathrm{ic} \gtrsim R_\nu$ is achieved, and the degree of
the deformation becomes larger for more rapidly rotating models.  In the following, we explain this
feature in more detail.

\begin{figure}[p]
    \centering
    \includegraphics[width=0.9\linewidth]{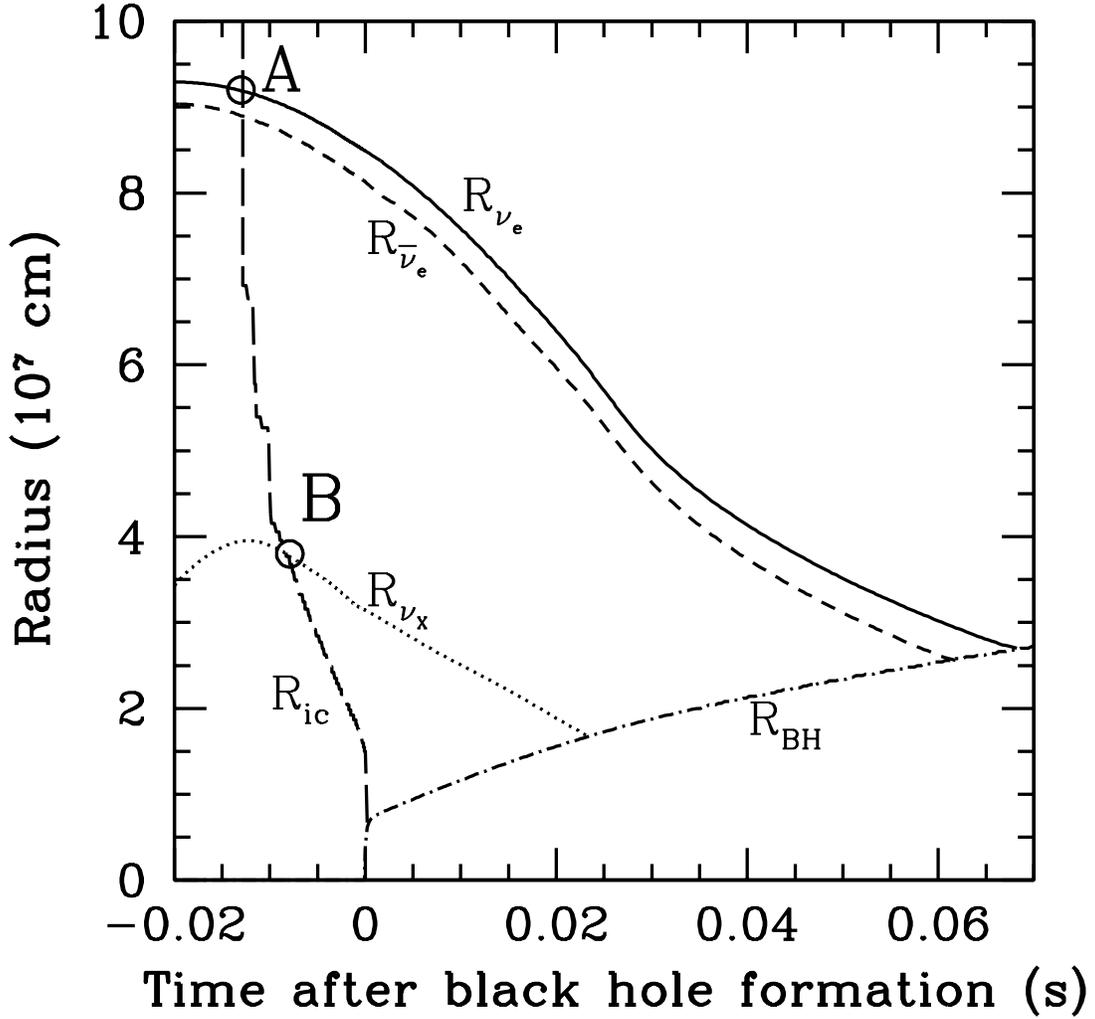}
    \caption{Radii of neutrinospheres of $\nu_e$ (solid), $\bar\nu_e$ (short-dashed), $\nu_X$
      (dotted), inner core surface (long-dashed) and BH (dot-dashed) versus time after black hole
      formation in seconds for the non-rotating model. The radii of the $\nu_e$ and $\bar\nu_e$
      neutrinospheres are similar because the electron fraction does not reach small value as in the
      case of core-collapse supernova. The inner core is inside all of the neutrinospheres just
      before BH formation so that the luminosity of the non-rotating model does not get large in
      Figure \ref{fig:neu_lumi}.}
    \label{fig:ns}
\end{figure}

\begin{figure}[p]
    \centering
    \includegraphics[width=1.\linewidth]{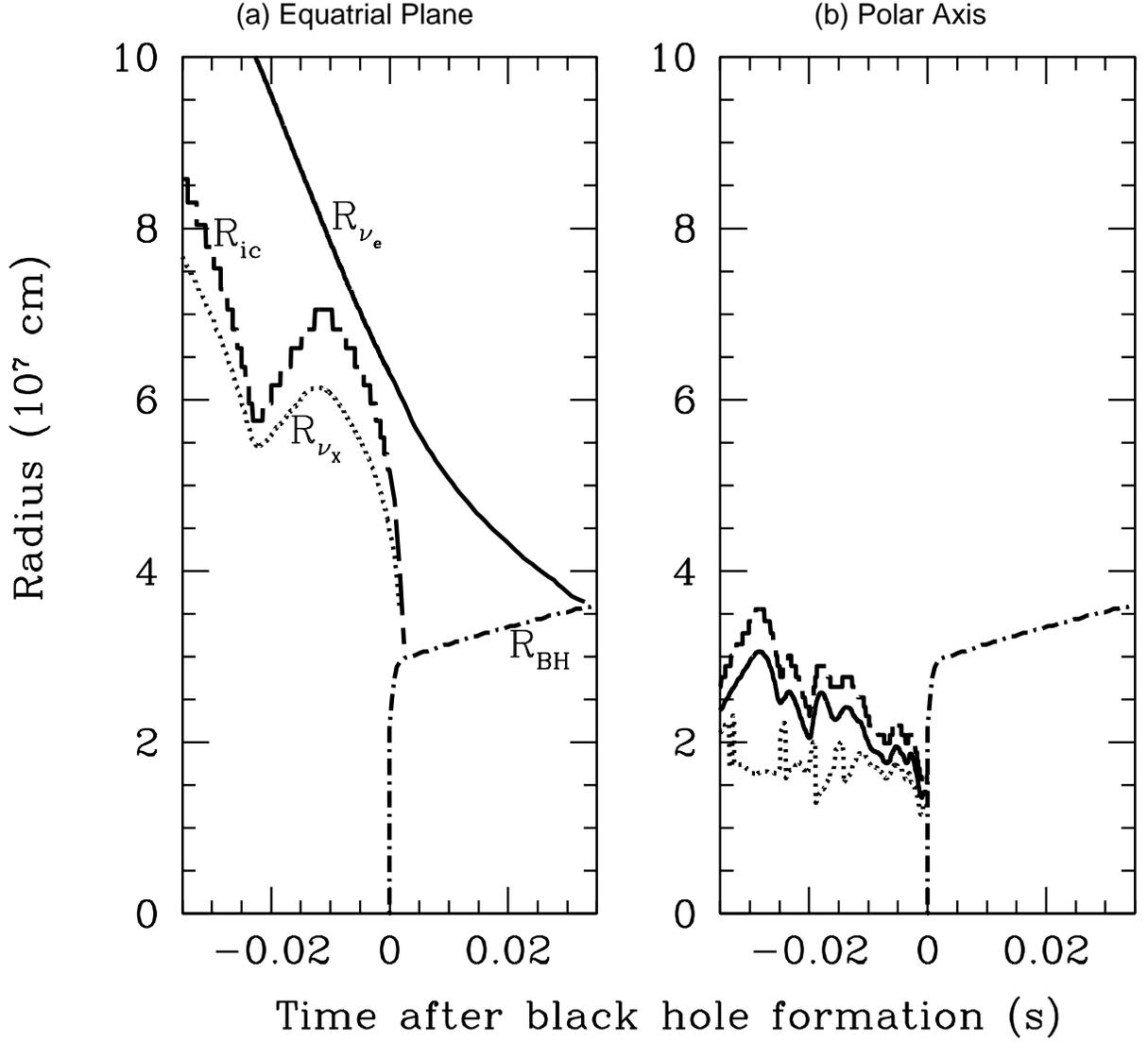}
    \caption{Time evolution of radii for a rotating model with $T/|W|_\mathrm{init}=0.5\%$. The left
      panel is in the equatorial plane and the right panel is along the polar axis.  The line types
      are the same with Figure \ref{fig:ns} except that the neutrinosphere of $\bar\nu_e$ is not
      shown in this figure because $R_{\bar\nu_e}\sim R_{\nu_e}$.  }
    \label{fig:neu_rot}
\end{figure}

\begin{figure}[p]
    \centering
    \includegraphics[width=0.45\linewidth]{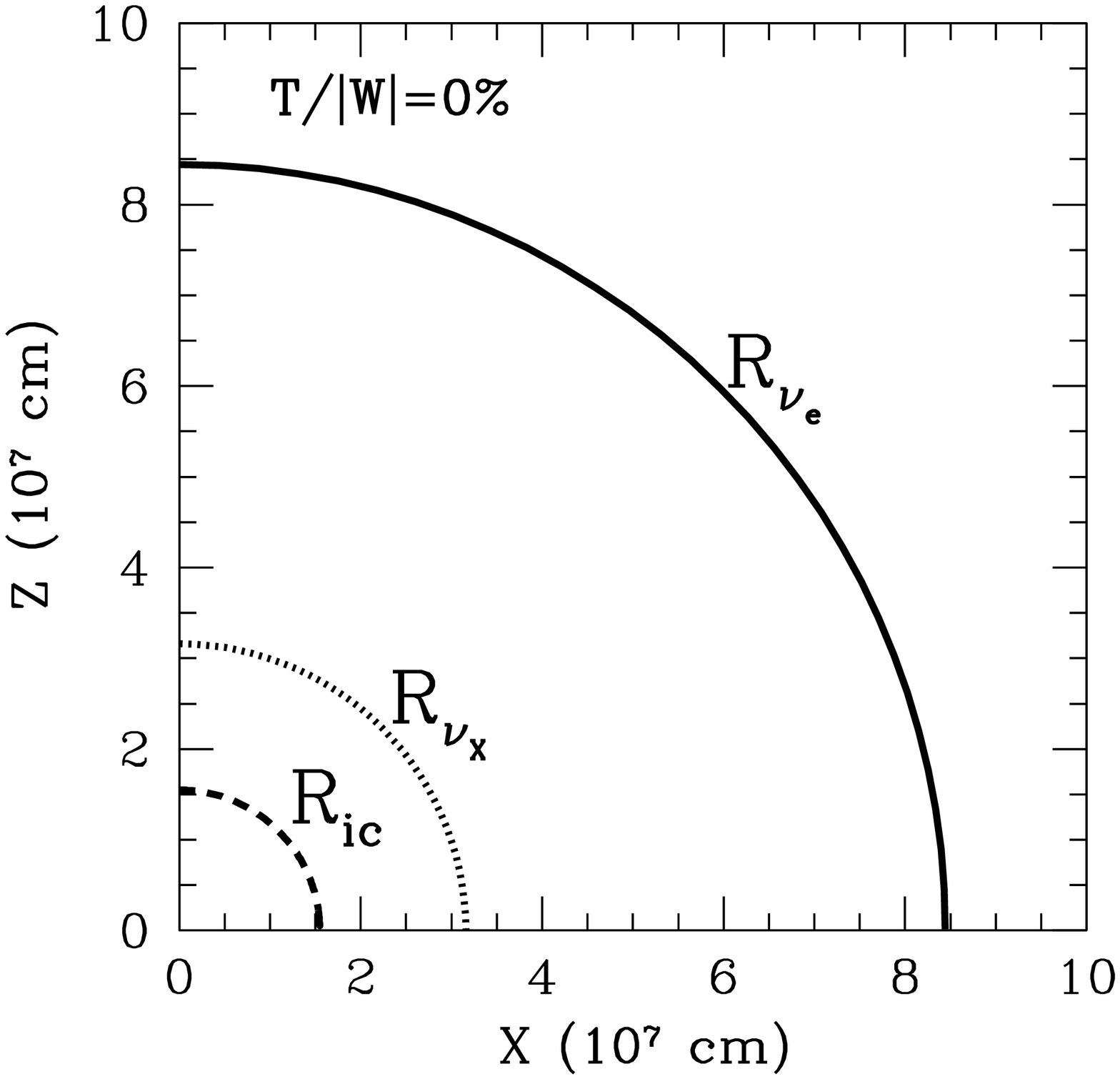}
    \includegraphics[width=0.45\linewidth]{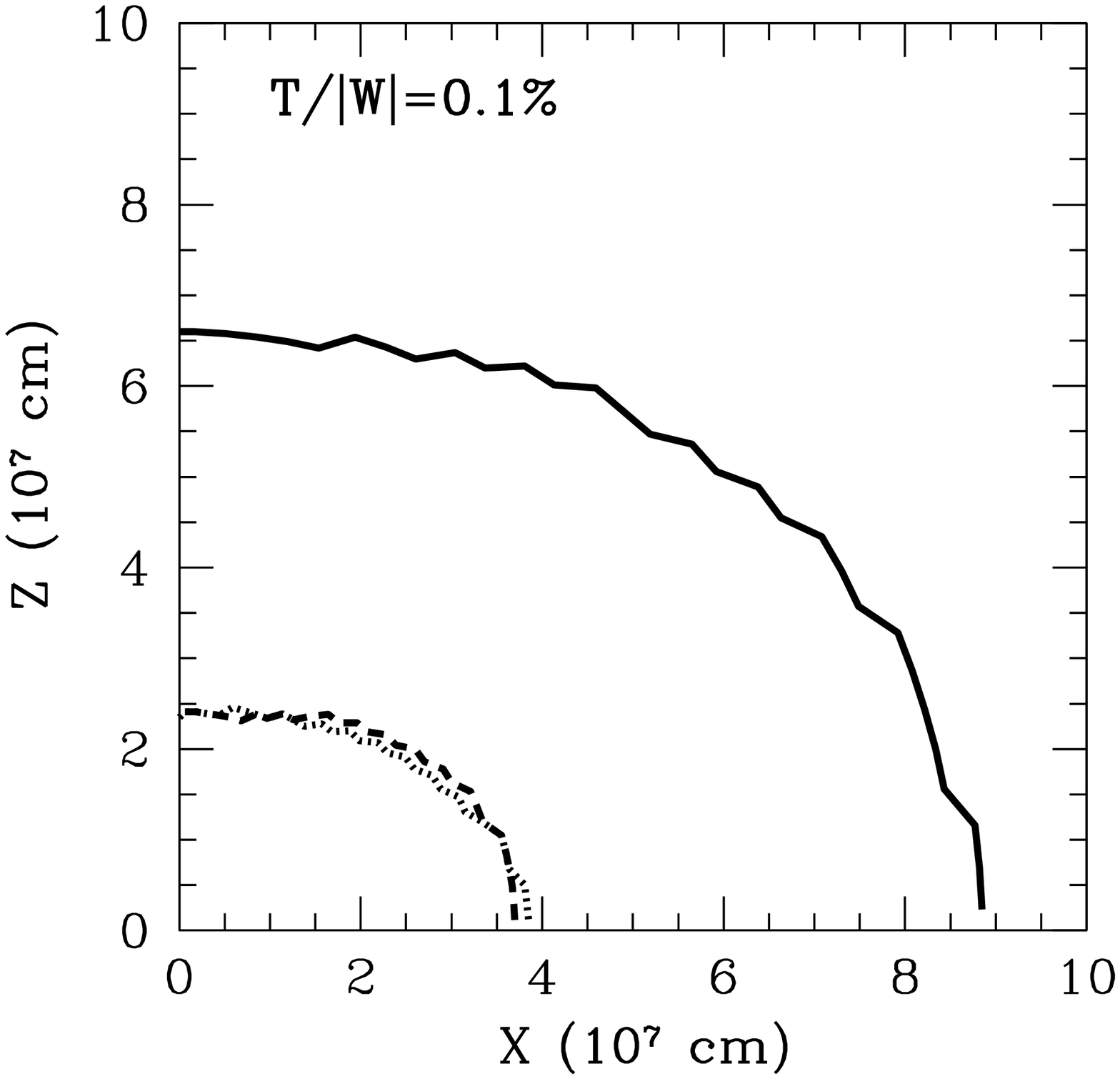}
    \includegraphics[width=0.45\linewidth]{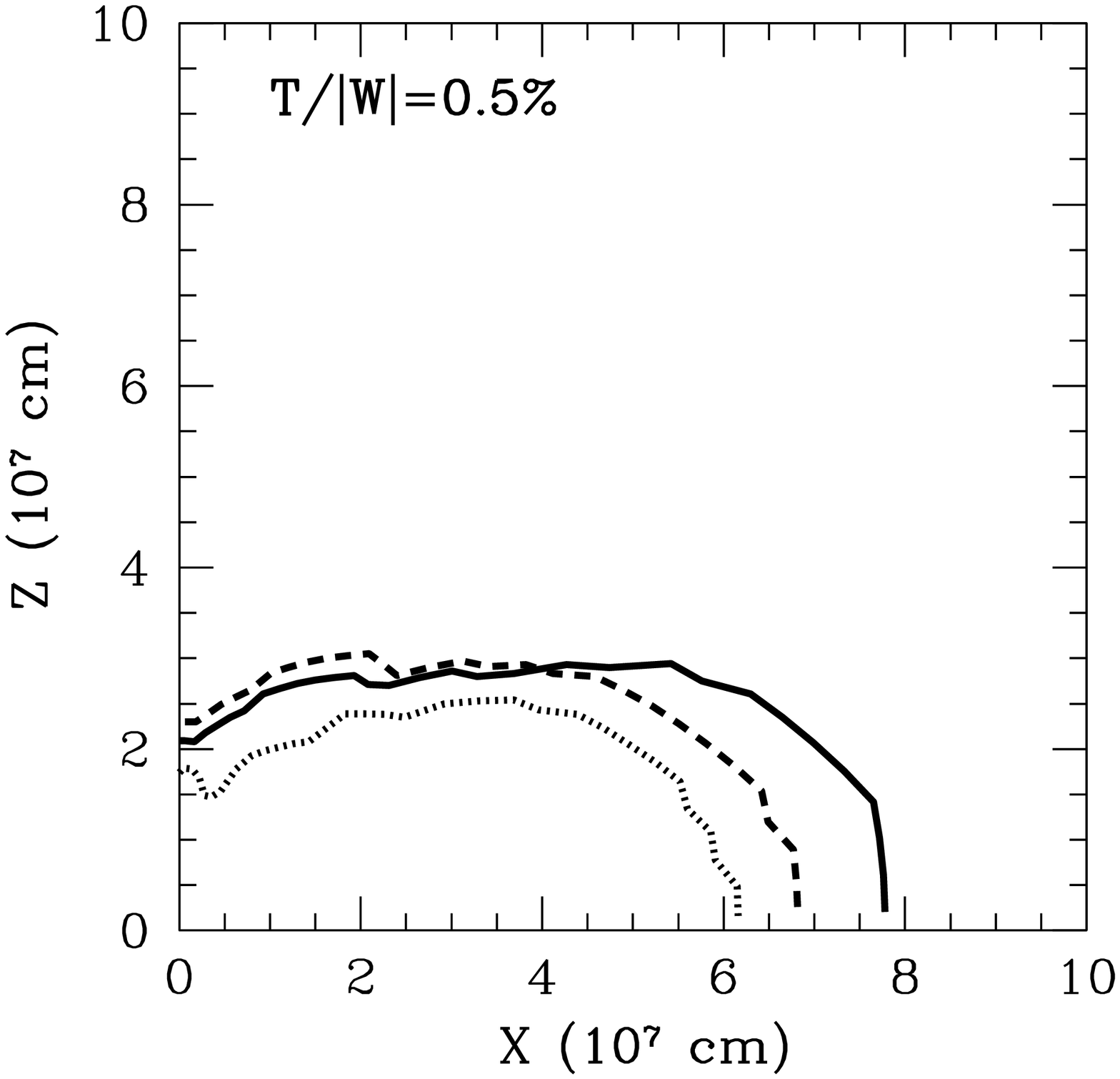}
    \caption{Positions of neutrinospheres and the inner core. The first panel shows the non-rotating
      model, the second $T/|W|_\mathrm{init}=0.1\%$ and the third $T/|W|_\mathrm{init}=0.5\%$. The
      solid, dotted and dashed lines give the locus of the neutrinospheres of $\nu_e$, $\nu_X$ and
      the inner cores, respectively.}
    \label{fig:ic_ns}
\end{figure}

Figures \ref{fig:ns} and \ref{fig:neu_rot} show the time evolution of radii of neutrinospheres and
inner core for non-rotating model and rotating model ($T/|W|_\mathrm{init}=0.5\%$), respectively.
Note again that we set the model with $T/|W|_\mathrm{init} = 0.5\%$ to be the reference rotating
model according to \citet{hege00,hege03a}.  For the non-rotating model (Figure \ref{fig:ns}), the
inner core becomes smaller than the surfaces of all of the neutrinospheres just before BH formation
(see evolutions after B in Figure\ref{fig:ns}).  On the other hand, for the rotating model, the
inner core is broadened in the equatorial plane due to the centrifugal forces (see left panel of
Figure \ref{fig:neu_rot}) and exists outside of the neutrinosphere of $\nu_X$.  Moreover, the inner
core surface is located outside of the neutrinosphere of $\nu_e$ along the polar axis (right panel).
The above effects make the neutrino luminosities larger for the rotating models.  In Figure
\ref{fig:ic_ns}, the global shapes of neutrinospheres of $\nu_e$ and $\nu_X$, and the inner cores
for spherical model and the rotating models with $T/|W|_\mathrm{init} = 0.1\%$, and $0.5\%$ are
presented.  It is clearly seen from the comparison with the non-rotating model that neutrinospheres
are more deformed for more rapidly rotating models.  As the initial rotation rate becomes larger,
the deformed inner core is shown to have larger surfaces approaching the neutrinospheres than the
non-rotating model.  As for the model with $T/|W|_\mathrm{init}=0.5\%$, the inner core extends
outside the neutrinosphere of $\nu_X$ entirely.

\begin{figure*}[p]
    \centering
    \includegraphics[width=1.\linewidth]{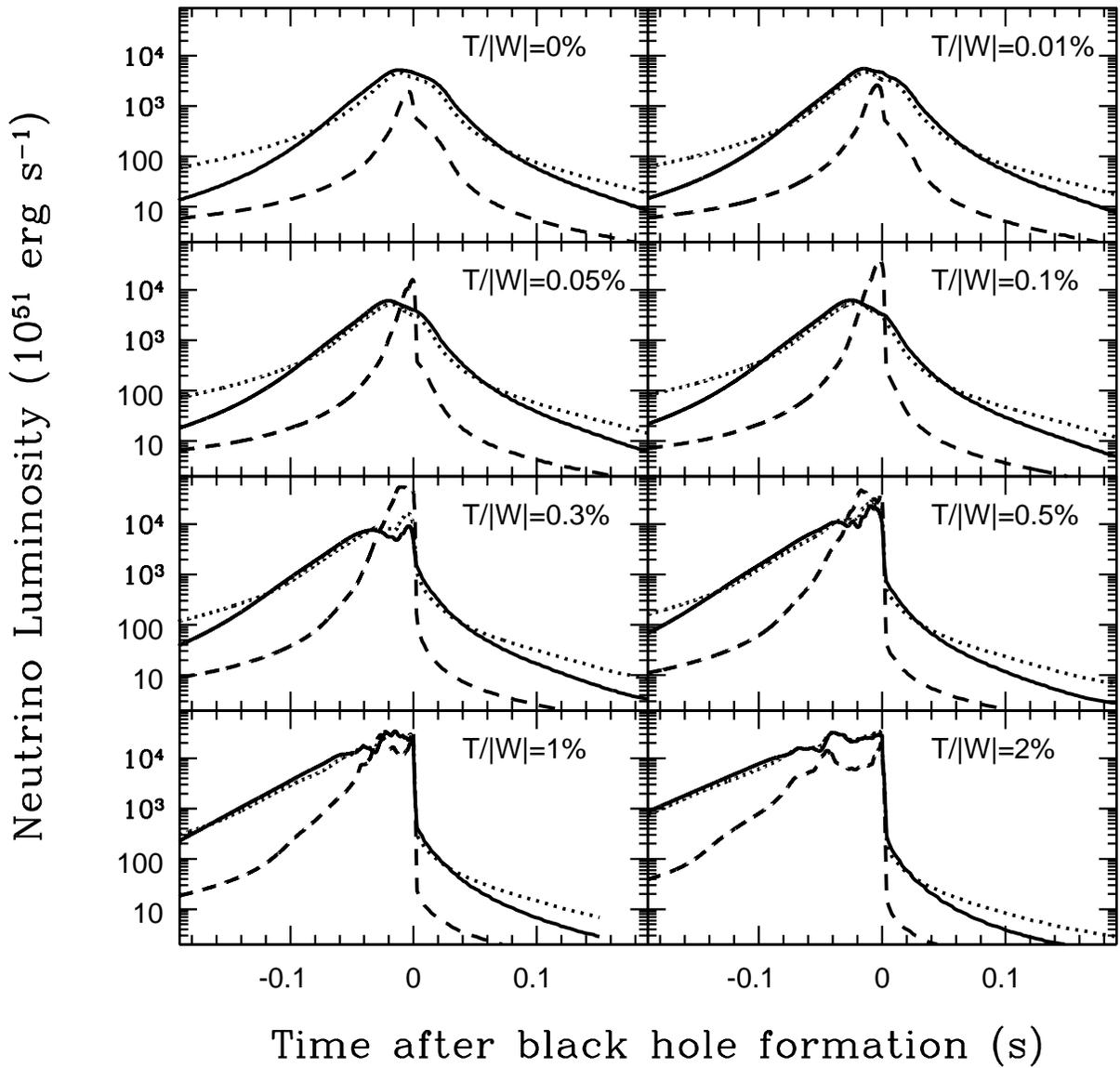}
    \caption{Time evolution of neutrino luminosity for all models.  $\nu_e$, $\bar\nu_e$ and $\nu_X$
      are represented by solid, dotted and dashed lines, respectively.}
    \label{fig:neu_lumi}
\end{figure*}

The time evolution of the neutrino luminosity for models with different initial strength of rotation
are shown in Figure \ref{fig:neu_lumi}.
For slow rotation models ($T/|W|_\mathrm{init}\lesssim 0.1\%$), the luminosity properties of $\nu_e$
and $\bar\nu_e$ are shown not to depend on the strength of rotation. On the other hand, the
luminosity of $\nu_X$ gets larger as the initial rotation gets faster.  This is because the radius
of the neutrinosphere of $\nu_X$ is smaller than those of $\nu_e$ and $\bar\nu_e$ so that $\nu_X$ is
most sensitive to the rotation.  The inner core surface, which is in the neutrinospheres for the
non-rotating model, is enlarged by rotation and falls close to the neutrinosphere of $\nu_X$ on the
equatorial plane, which leads to the increase of $\nu_X$ luminosity.
In fast rotation models ($T/|W|_\mathrm{init}\gtrsim0.3\%$), the luminosity of $\nu_e$ and
$\bar\nu_e$ also get larger with rotation for the same reasons as for the $\nu_X$ case.  The
luminosity of $\nu_X$, however, gets smaller as the rotation gets faster.  This feature was already
seen in \cite{suwa07a}.  This is because rapid rotation suppress the contraction of the most inner
part causing the density and temperature not to increase as much as in the case of slower rotation.
Therefore, the emissivity of neutrinos also does not increase very much.  This suppression affects
the $\nu_e$ and $\bar\nu_e$ emission less because neutrinospheres of these species are located at
larger radii.

From Figure \ref{fig:neu_lumi}, we can see drastic decreases in neutrino luminosities near the epoch
of black hole formation, particularly for the rapidly rotating models.  The slope of the neutrino
luminosity depends on the relative position of the neutrinospheres and the inner core.  For slow
rotation models, only $\nu_X$ shows such a decrement because the neutrinosphere of $\nu_X$ is the
smallest so that it is absorbed first and is inside of the BH when the BH forms. On the other hand,
the neutrinospheres of $\nu_e$ and $\bar\nu_e$ remain outside of the BH at the onset of BH formation
and are absorbed as the BH grows. The luminosity of $\nu_e$ and $\bar\nu_e$ decrease with the
dynamical timescale ($\sim O(10)$ ms) of the region of the neutrinospheres. For fast rotation
models, all species show the drastic decrease because all neutrinospheres are partly or entirely in
the BH region when it forms (see Figure \ref{fig:neu_rot}, again) because the rapid rotation leads
to a larger initial mass of the BH \cite[see][]{suwa07a}.

We find that rotation enhances the total energy released by neutrinos.  The total energy emitted in
the form of neutrinos during the collapse is summarized in Table \ref{tab}.  It can be seen that for
the non-rotating model the total radiated energy amounts to $6.6 \times 10^{53}$ ergs.  This result
is in good agreement with the simulation of \cite{naka06}, who included neutrino transport during
the collapse of a 300 $M_\odot$ Pop III star. They found a total energy output of $\sim 4 \times
10^{53}$ ergs.  The stronger rotation models show larger energy emission. This is because the
rotation enhances the neutrino luminosity as already described. In addition the rotation delays BH
formation by the centrifugal force, which leads to a longer duration of neutrino emission. These two
effects account for the increase of radiated energy by neutrinos as the strength of rotation
increases.

\subsection{Features of Neutrino Energy} \label{sec:ene}

\begin{figure}[p]
    \centering
    \includegraphics[width=1.\linewidth]{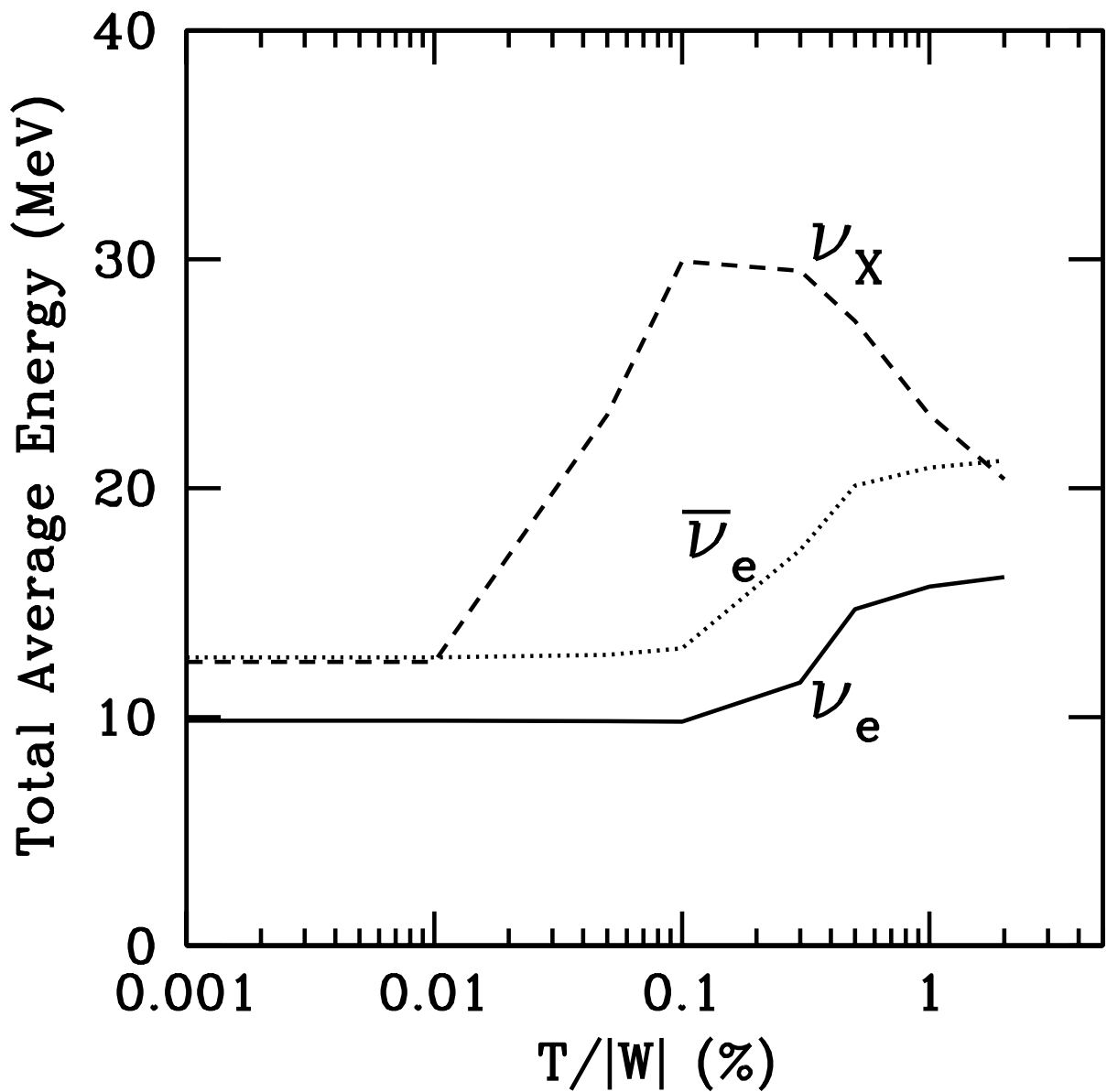}
    \caption{The total average energy of neutrinos, $\bracket{E_\nu}\equiv \int Q_\nu dV dt/ \int
      N_\nu dV$, where $N_\nu$ is the number density of emitted neutrinos, as a function of
      $T/|W|_\mathrm{init}$.  The solid, dotted and dashed lines represent $\nu_e$, $\bar\nu_e$ and
      $\nu_X$, respectively.}
    \label{fig:average}
\end{figure}

We discuss the effects of rotation on the emitted neutrino energies in this subsection. We consider
the averaged neutrino energy, $\bracket{E_\nu}\equiv \int Q_\nu dV dt/ \int N_\nu dV$, where $N_\nu$
is number density of emitted neutrinos.  Figure \ref{fig:average} shows the difference of the
average energy as a function of $T/|W|_\mathrm{init}$.  It can be seen that the average energies
increase with the initial rotation rate for $T/|W|_\mathrm{init}\lesssim 0.1\%$ irrespective of the
flavors, which is basically the same reason as discussed in the previous subsection.  It should be
noted that the energy of $\nu_X$ is more sensitive to the rotation strength than the other species
because the neutrinosphere of $\nu_X$ forms the deepest in the core.  In fact, only $\nu_X$ is
affected for relatively slow rotation ($T/|W|_\mathrm{init}\lesssim 0.1\%$). As the rotation becomes
stronger, the energies of $\nu_e$ and $\bar\nu_e$ are gradually changed.  Although the average
energies of neutrinos basically follow the standard order sequence, namely
$\bracket{E_{\nu_e}}<\bracket{E_{\bar\nu_e}}<\bracket{E_{\nu_X}}$ as described in \citet{iocco05},
$\bracket{E_{\nu_X}}$ is found to be smaller than $\bracket{E_{\bar\nu_e}}$ for the model with
$T/|W|_\mathrm{init}=2\%$. The strong centrifugal force interrupts the contraction of the inner part
of the core, leading to the suppression of the releasable gravitational energy and thus smaller
energy of neutrinos.

\begin{figure*}[p]
    \centering
    \includegraphics[width=\linewidth]{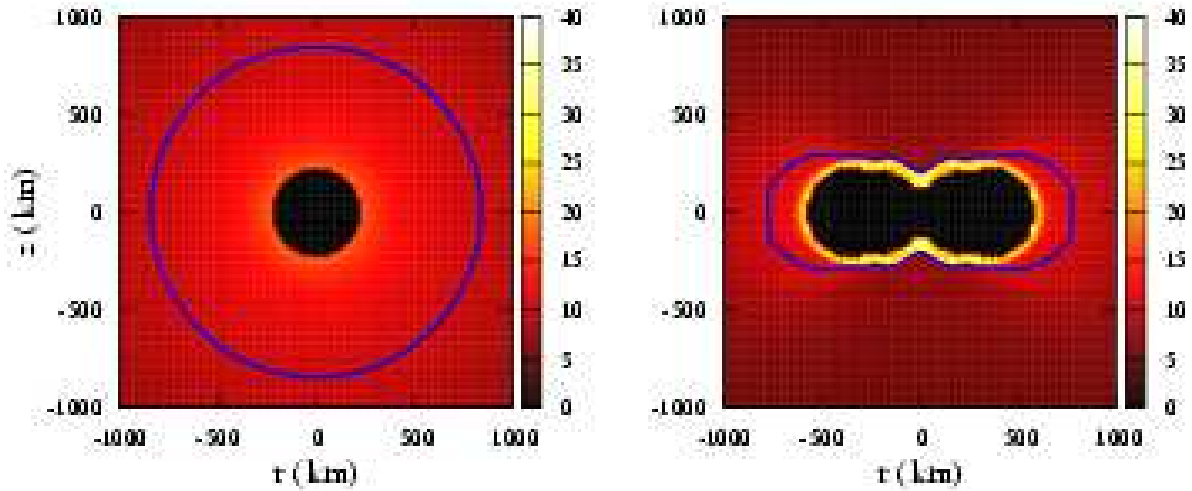}
    \caption{ Average energy distribution of $\nu_e$ and the shape of neutrinosphere just before BH
      formation for models with $T/|W|_\mathrm{init}=0\%$ (left) and $0.5\%$ (right).  The blue
      solid line represents the position of the neutrinosphere. The central black part corresponds
      to the region with the optical depth of neutrino being beyond 10, where neutrinos are almost
      trapped.}
    \label{fig:neu_dist}
\end{figure*}

Next, we focus on the anisotropy of the neutrino energies and emission.  The spatial distribution of
the average energy of $\nu_e$ just before BH formation, with and without rotation are depicted in
Figure \ref{fig:neu_dist}. It can be seen that due to the rotational flattening of the
neutrinosphere (blue line) near the polar regions (bottom panel), the neutrinos achieve the higher
temperature just inside the neutrinosphere, leading to the emission of higher energy neutrinos. In
addition, the direction of anisotropy of the energies can be seen in the rotating model. In the
polar direction, the inner core surface is almost coincident with the neutrinosphere (see Figure
\ref{fig:neu_rot} and \ref{fig:ic_ns}, again) so that neutrinos with high energy produced just
inside the inner core can escape. On the other hand, the optical depth is $\sim 5$ at the inner core
on the equatorial plane and few neutrino with high energy can escape.  These effects, irrespective
of the neutrino flavors, make the neutrino spectrum of the rotating model broader and harder than
the non-rotating model, which will be useful for the discussion in the following subsection.

\subsection{Neutrino Spectrum and Relic Background} \label{sec:spe}

\begin{figure}[p]
    \centering
    \includegraphics[width=1.\linewidth]{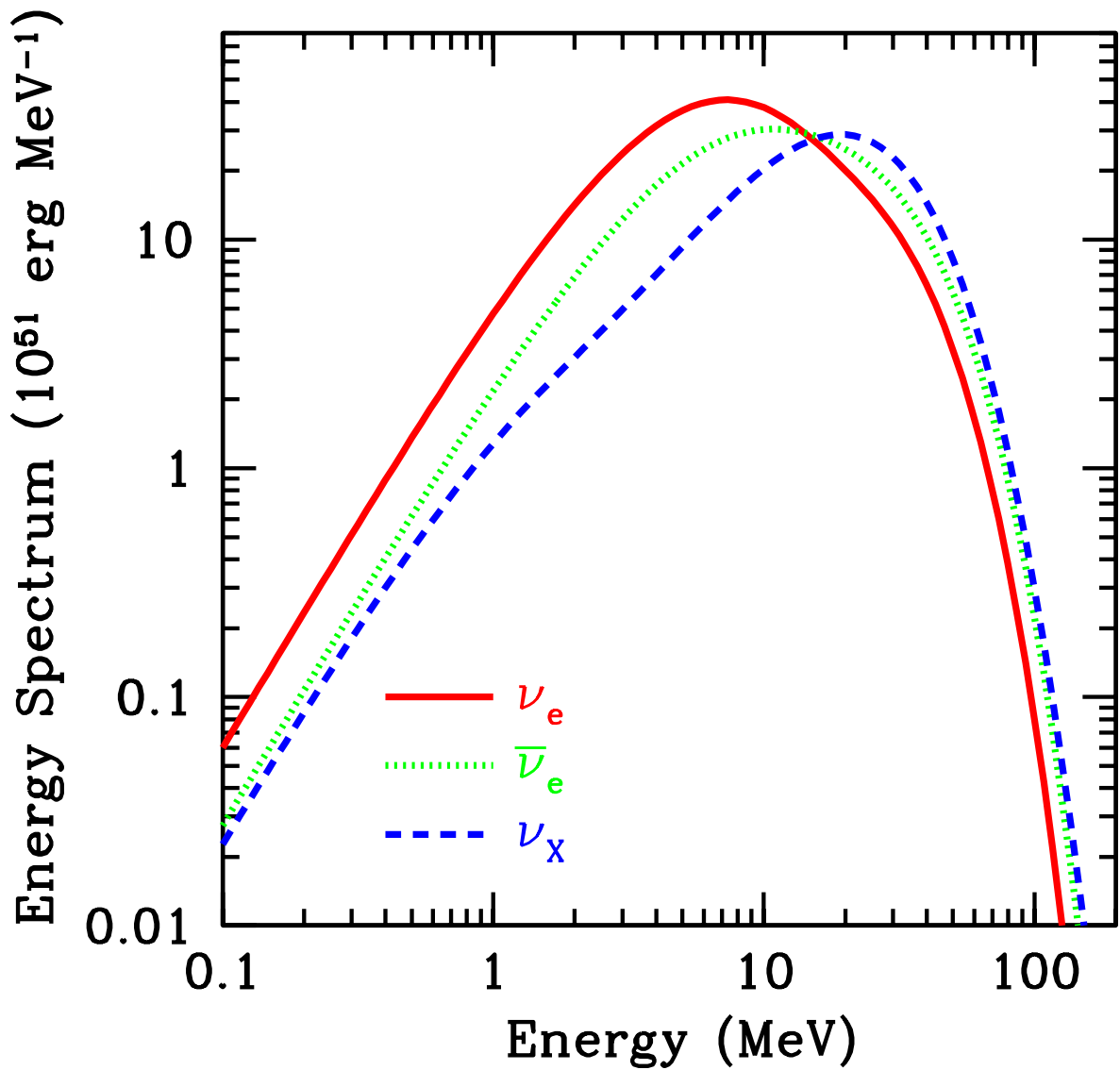}
    \caption{Energy spectra of neutrinos for the model with $T/|W|_\mathrm{init}=0.5\%$. The red
      solid, green dotted and blue dashed lines represent $\nu_e$, $\bar\nu_e$ and $\nu_X$,
      respectively.}
    \label{fig:diff_spec}
\end{figure}

\begin{figure}[p]
    \centering
    \includegraphics[width=1.\linewidth]{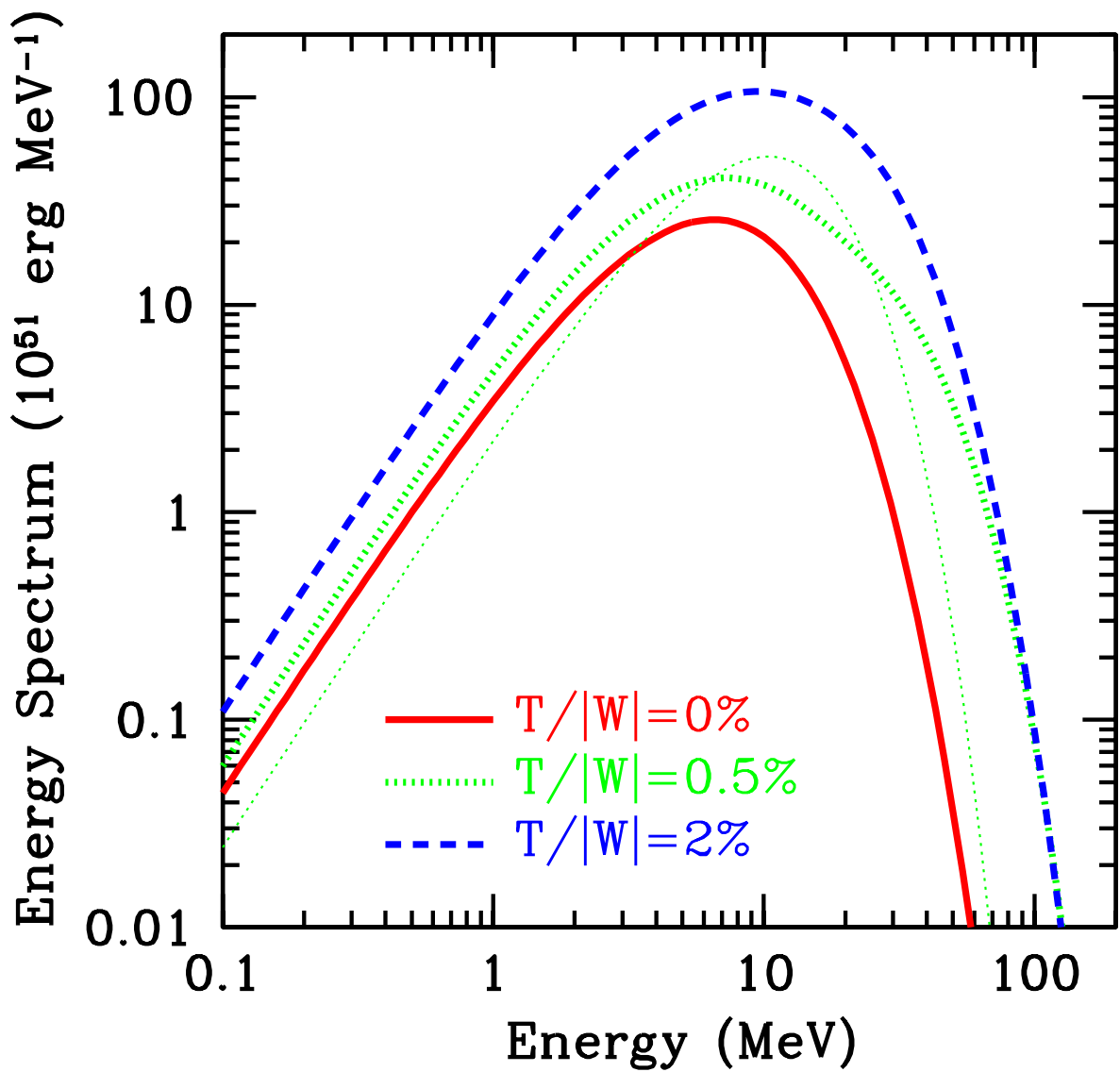}
    \caption{Energy spectrum of $\nu_e$ for model with different values of $T/|W|_\mathrm{init}$.
      The red solid, green thick-dotted and blue dashed lines represent $T/|W|_\mathrm{init}=0$, 0.5
      and 2\%, respectively.  The green thin-dotted line shows the spectrum with the single
      temperature Fermi-Dirac distribution function (see \S\ref{sec:spe} for discussion).}
    \label{fig:spec}
\end{figure}

In this subsection, we discuss the features of the neutrino spectrum and the effects of rotation.
We calculate the neutrino luminosity spectrum as follows:
\begin{equation}
    \frac{dL_{\nu_\alpha}}{d\epsilon'}(t)=\int dV Q_{\nu_\alpha}\frac{dP_{\nu_\alpha}}{d\epsilon'},
    \label{eq:dlde}
\end{equation}
where $Q_{\nu_\alpha}$ is the emissivity for $\nu_\alpha$ ($\nu_\alpha=\nu_e, \bar\nu_e,$ and
$\nu_X$) and $dP_{\nu_\alpha}/d\epsilon$ is the normalized Fermi-Dirac distribution,
\begin{equation}
    \frac{dP_{\nu_\alpha}}{d\epsilon'}=\frac{2}{3\zeta_3 T_\alpha^3}\frac{\epsilon'^2}{\e^{\epsilon'/T_\alpha}+1},
    \label{eq:dpde}
\end{equation}
where $\epsilon'$ is the energy of emitted neutrinos in the source frame and
$T_\alpha=180\zeta_3\bracket{\epsilon'_{\nu_\alpha}}/7\pi^4$ is the effective neutrino temperature
with $\bracket{\epsilon'_{\nu_\alpha}}$ being the local average energy of $\nu_\alpha$.  Figure
\ref{fig:diff_spec} shows the time integrated energy spectrum, $\int dt
(dL_{\nu_\alpha}/d\epsilon')$, of $\nu_e$, $\bar\nu_e$ and $\nu_X$ for the model with
$T/|W|_\mathrm{init}=0.5\%$. It can be seen that the spectrum of $\nu_X$ (dashed line) is harder
than $\nu_e$ (solid line) and $\bar\nu_e$ (dotted line) as already described in the previous
subsection.  Figure \ref{fig:spec} shows time integrated energy spectrum of $\nu_e$ for models with
$T/|W|_\mathrm{init}=0, 0.5$ and $2\%$. It is obvious that the spectra of rotating models (dashed
and thick dotted lines) are higher and harder than non-rotating model (solid line). In this figure,
the thin dotted line represents the single energy spectrum.  We use the total energy emitted by
$\nu_e$, $\int L_{\nu_e}dt$, and total average energy, $\bracket{E_{\nu_e}}$, of the model with
$T/|W|_\mathrm{init}=0.5\%$ for this spectrum to compare the spectrum of the model with the same
rotation strength. By making the comparison between thin and thick dotted lines, it can be seen that
the calculated spectrum (thick dotted line) is harder than one with a single energy. This is because
the neutrino temperature is not a simple, single temperature component, but rather consists of
multiple temperature components, depending on the time and position.  It should be noted that the
time evolution of the average energy from Pop III stars is drastic unlike the ordinary core-collapse
supernovae, which do not show significant change of the average energy on the cooling phase of
proto-neutron stars \citep[see, e.g.,][]{lieb05}, so that the spectrum becomes broader than the
ordinary supernovae.  The single energy spectrum employed in previous works \citep{iocco05,daig05}
should have led to underestimates in the neutrino number in the high energy regime.

\begin{figure}[p]
    \centering
    \includegraphics[width=1.\linewidth]{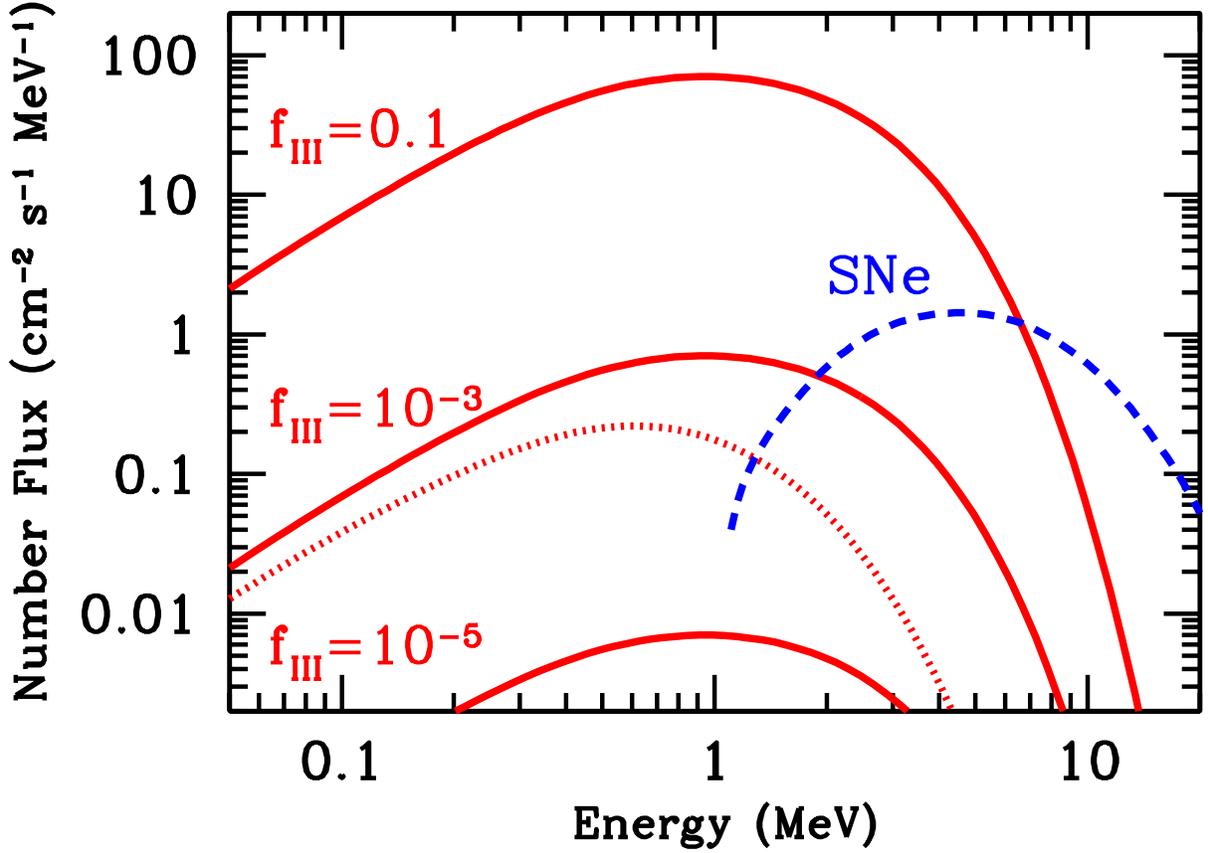}
    \caption{Calculated number flux of $\bar\nu_e$ (red solid lines). The employed baryon fraction
      of Pop III stars, $f_{III}$, is $10^{-1}, 10^{-3}$ and $10^{-5}$ from top to bottom. As for
      the solid lines, we employed the result of the $T/|W|_\mathrm{init}=2\%$ model. The dotted
      line represents the flux with the $T/|W|_\mathrm{init}=0\%$ model and $f_{III}=10^{-3}$.  For
      comparison we show the diffuse flux from ordinary core-collapse supernovae (blue dashed line)
      according to \citet{ando03}.}
    \label{fig:relic}
\end{figure}

We are now in a position to discuss the relic neutrino background from Population III stars.  For
simplicity, we assume that all Pop III have identical emission characteristics. According to
\citet{iocco05} the differential flux expected at Earth can be written as
\begin{eqnarray}
    \frac{dF_{\nu_\alpha}}{d\epsilon}=cf_{III} n_\gamma\eta\frac{m_N}{M_{III}}\int^\infty_0 dz (1+z) 
    \psi(z) \frac{dN_{\nu_\alpha}}{d\epsilon'},\label{eq:dfde}
\end{eqnarray}
where $\epsilon=\epsilon'/(1+z)$ is the observed energy of neutrinos, $f_{III}$ is the fraction of
all baryons going through Pop III, $n_\gamma\simeq 410 \mathrm{cm}^{-3}$ is the cosmic microwave
background photon density at redshift zero, $\eta\simeq 6.3\times 10^{-10}$ is the cosmic
baryon-photon ratio, $m_N$ is the nucleon mass, $M_{III}$ is the mass of Pop III stars, $\psi(z)$ is
normalized star formation rate, and $dN_{\nu_\alpha}/d\epsilon'$ represents the number spectra at
the source frame, which is also calculated by volume integration with the normalized Fermi-Dirac
distribution (Eq.  (\ref{eq:dpde})).  We assume that the star formation rate is concentrated around
a redshift $z\sim 10$, that is, $\psi(z)=\delta(z-10)$. In addition, we assume that all Pop III
stars have the same mass, $M_{III}=300M_\odot$.  We used the result of the $T/|W|_\mathrm{init}=2\%$
model to calculate the relic neutrino flux.  Since star formation history is not well constrained
observationally at high z, we consider samples of $f_{III}$ as $10^{-1}$, $10^{-3}$, and $10^{-5}$.
Figure \ref{fig:relic} shows the calculated number flux spectrum of $\bar\nu_e$.  If we adopt the
largest value for $f_{III}$, the diffuse anti-electron neutrinos from Pop III dominates those from
ordinary supernovae (dashed line) below $\sim$ 7 MeV. The dotted line represents the flux from the
model with $T/|W|_\mathrm{init}=0\%$ and $f_{III}=10^{-3}$. Comparing with the second solid line
from the top, we can see the amplification of the relic neutrino flux by the stellar rotation.

\section{Summary and Discussion}

We performed a series of 2D hydrodynamic simulations of rotational collapse of Pop III stars. We
computed 8 models, changing the initial strength of the angular momentum in a parametric manner. We
studied systematically the deformation of neutrinospheres and investigated the degree of anisotropic
neutrino emission from the point of view of the neutrino luminosity and the average energy, and
their impacts on the neutrino spectrum. Then we found the following,
\begin{enumerate}
\item The neutrino luminosity increases for more rapidly rotating cores. This is because the inner
    core, which is the dominant neutrino source, is deformed due to the rotation and can extend
    outside the neutrinospheres.  In particular, $\nu_X$ is most sensitive to the rotation because
    its neutrinosphere is located at the smallest radius and nearest to the inner core.  The total
    energy emitted by neutrinos also increases with the strength of rotation.
\item The average energy of emitted neutrinos gets larger as rotation becomes faster. In addition,
    the energy is larger in the polar direction than in the equatorial plane due to the rotation.
    The neutrino spectrum gets broader and harder for rotating models. As a result, the number
    spectrum of the relic neutrinos also gets harder so that the relic neutrinos from Pop III stars
    overwhelm those from ordinary core-collapse supernova below $\sim 7$ MeV, provided a fraction
    $f_{III}=0.1$ of all baryons forms Pop III stars.
\end{enumerate}

As for our numerical computations, we shall address a few of our assumptions.  In this study, we
mimicked the neutrino transfer by the leakage scheme.  Although the scheme is a radical
simplification, we checked that in the case of spherical collapse, the obtained integral quantities
such as the total energy emitted by neutrinos and their average energy are very close to the results
by \citet{naka06}, in which the Boltzmann neutrino transport was solved.  We ignored the angle
dependence of the neutrino emission, assuming that the neutrino track is radial. This approximation
will break down just before BH formation because the neutrinosphere becomes oblate and the neutrino
emission would be highly aspherical. Nonetheless, the multi-angle and multi-dimensional neutrino
transfer is still computationally prohibitive \citep[recently reported by][in the framework of
core-collapse supernova]{ott08}.  The angle dependence is expected to make the neutrino spectrum
harder. This is because the neutrinos produced in the vicinity of the equatorial plane might be able
to escape to the polar direction even if they cannot escape radially, thus it should be taken into
account.  Furthermore the Newtonian simulation in this paper, albeit with the general relativistic
corrections, is nothing but an idealized study to describe the dynamics with BH formation. It is by
no means definitive.  In order to explore these phenomena in more detail will require fully general
relativistic simulations \citep[e.g.,][for BH formation of massive stars]{shib03,seki05,liu07} but
with the microphysical ingredients at least at the same level of this study, which is beyond the
scope of this paper.

Bearing these caveats in mind, we state some discussions and speculations about the relic neutrinos
from Pop III stars.  In estimating the relic neutrino background, we employed a parameter $f_{III}$,
the baryon fraction of Pop III stars.  Because this parameter is entirely unknown so far, we
employed three values of $10^{-1}$, $10^{-3}$, and $10^{-5}$.  The largest value is obtained for the
explanation of the infrared background excess by UV photons from Pop III stars
\citep{sant02,dwek05}. So our estimation should be interpreted as giving an upper bound.  As for the
star formation history, we employed only the delta function one because the baryon fraction
dominates the uncertainty and the other ones such as the evolution of star formation rate are minor
corrections \citep[see,][for the discussion of the dependence on the star formation
history]{iocco05,daig05,naka06}.
In addition, we assume that all Pop III stars have the same mass of $300 M_\odot$ for simplicity.
The dependence of the progenitor masses should be also clarified, although the qualitative features
induced by rotation studied here will not change.

The predicted foregrounds for the relic neutrinos in the relevant energy range computed here are the
neutrinos produced in the Sun and the anti-neutrinos from nuclear reactors.  These (anti-)neutrinos,
however, could be removable in principle because the distribution of these foregrounds are not
isotropic.  Other possible foregrounds, which are isotropic, are extra-galactic {\it neutrinos}
produced by thermonuclear reactions in stars \citep{porc04}.  On the other hand, as for {\it
  anti-neutrinos} the contribution of Pop III stars are dominant in the range below a few MeV.  Thus
we speculate that the detection of this signature could be the first direct probe of this as yet
unseen metal-free first generation of stars.

\acknowledgments

We would like to thank Shin'ichiro Ando for fruitful comments and Erik Reese for reading manuscript.
Numerical computations were in part carried on XT4 and general common use computer system at the
center for Computational Astrophysics, CfCA, the National Astronomical Observatory of Japan.  This
study was supported in part by the Japan Society for Promotion of Science (JSPS) Research
Fellowships (Y.S.), Grants-in-Aid for the Scientific Research from the Ministry of Education,
Science and Culture of Japan (No.S19104006, No.20740150).

\bibliography{bib}

\begin{table}[p]
    \centering
    \caption{Results of the numerical simulations.}
    \begin{tabular}{ccc}
        \hline\hline
        $T/|W|_\mathrm{init}$ & $L_\nu^\mathrm{peak}$ & $E_\mathrm{total}$\\
        (\%)                  & ($10^{54}$ erg/s)     &  ($10^{53}$ erg)  \\
        \hline
        0    & 1.11 & 6.57  \\
        0.01 & 1.17 & 6.78  \\
        0.05 & 2.17 & 8.30  \\
        0.1  & 4.19 & 11.3  \\
        0.3  & 7.55 & 22.2  \\
        0.5  & 8.22 & 29.6  \\
        1    & 8.64 & 40.8  \\
        2    & 8.44 & 52.4  \\
        \hline\hline
    \end{tabular}
    \tablecomments{
      $T/|W|_\mathrm{init}$ means the initial ratio of the rotational energy, $T$, and the 
      gravitational energy, $W$. $L_\nu^\mathrm{peak}$ represents the peak total luminosity of 
      neutrinos. $E_\mathrm{total}$ is the total energy emitted by neutrinos.
    }
    \label{tab}
\end{table}

\end{document}